\documentstyle[12pt]{article}
\begin{document}\begin{flushright}\thispagestyle{empty}
OUT--4102--74\\
hep-th/9806174\\
1 July 1998                    \end{flushright}\vspace*{2mm}\begin{center}{
                                                       \Large\bf
Solving differential equations for 3-loop diagrams:    \\[5pt]
relation to hyperbolic geometry and knot theory        }\vglue 10mm
                                                       {\large{\bf
D.~J.~Broadhurst                                       $^{1)}$}\vglue 4mm
Physics Department, Open University                    \\[3pt]
Milton Keynes MK7 6AA, UK     }\end{center}\vfill\noindent{\bf Abstract}\quad
In hep-th/9805025, a result for the symmetric 3-loop massive tetrahedron in 3
dimensions was found, using the lattice algorithm PSLQ. Here we give a more
general formula, involving 3 distinct masses. A proof is devised, though it
cannot be accounted as a derivation; rather it certifies that an Ansatz found
by PSLQ satisfies a more easily derived pair of partial differential
equations. The result is similar to Schl\"afli's formula for the volume of a
bi-rectangular hyperbolic tetrahedron, revealing a novel connection between
3-loop diagrams and 1-loop boxes. We show that each reduces to a common basis:
volumes of ideal tetrahedra, corresponding to 1-loop massless triangle
diagrams. Ideal tetrahedra are also obtained when evaluating the volume
complementary to a hyperbolic knot. In the case that the knot is positive, and
hence implicated in field theory, ease of ideal reduction correlates with
likely appearance in counterterms. Volumes of knots relevant to the number
content of multi-loop diagrams are evaluated; as the loop number goes to
infinity, we obtain the hyperbolic volume of a simple 1-loop box.
\vfill\footnoterule\noindent
$^1$) D.Broadhurst@open.ac.uk;
http://physics.open.ac.uk/$\;\widetilde{}$dbroadhu
\newcommand{\df}[2]{\mbox{$\frac{#1}{#2}$}}\def\bk{\mbox{\bf k}}
\newpage\setcounter{page}{1}

\section{Introduction}

In~\cite{CTet} we studied the 3-loop 3-dimensional tetrahedral Feynman
diagram
\begin{equation}
C(a,b):=\frac{1}{\pi^6}\int\int\int\frac{d^3\bk_1d^3\bk_2d^3\bk_3}
{(k_1^2+a^2)(k_2^2+1)(k_3^2+1)(k_{2,3}^2+b^2)(k_{1,3}^2+1)(k_{1,2}^2+1)}
\label{cab}
\end{equation}
with $k_n^2:=|\bk_n|^2$ and $k_{i,j}^2:=|\bk_i-\bk_j|^2$. For the
totally symmetric tetrahedron, with $a=b=1$,
we found a simple reduction to a Clausen integral:
\begin{equation}
{C(1,1)\over2^{5/2}}=
-\int_{2\alpha}^{4\alpha}d\theta\log(2\sin\df12\theta)\,;
\quad\alpha:=\arcsin\df13\,,\label{ans}
\end{equation}
thus obtaining an exact dilogarithmic result for the diagram
evaluated numerically in~\cite{GKM}.

The discovery route for~(\ref{ans}) was based on a dispersion relation
for the more general Feynman tetrahedron~(\ref{cab}),
with masses $a$ and $b$ on non-adjacent lines
and unit masses on the other 4 lines. This was derived by applying
the methods of~\cite{mas,sixth} in 3 dimensions.
In this paper, we reduce $C(a,b)$ to 7 dilogarithms, for $a^2+b^2>4$,
and to 8 Clausen values, for $a^2+b^2<4$. In the latter case,~(\ref{ans})
results by use of the classical formula~\cite{BBP}
\begin{equation}
\pi=2\arcsin\df13+4\arcsin\df{1}{\sqrt3}\,,\label{class}
\end{equation}
which reduces the Clausen values to only 2.

Section~2 gives the 3-loop results. In Sections~3~and~4
we examine connections, via hyperbolic geometry,
to very different types of diagrams, in 4 dimensions:
massive box diagrams, with only 1 loop, studied in~\cite{DD}, and
massless diagrams with more than 6 loops, studied
in~\cite{BK15,BGK}. Remarkably, the infinite-loop limit of the
hyperbolic volumes of knots entailed by the latter
recovers a simple case of the former. Section~5 gives our conclusions.

\section{Solving the vacuum differential equations}

In~\cite{CTet}, we reduced~(\ref{cab}) to dispersive
integrals of the form $\int dx\, P(x,X)\log Q(x,X)$
where $P$ and $Q$ are rational algebraic functions of $x$ and of the
square root, $X$, of a quadratic function of $x$.
Section~8.1.2 of~\cite{Lewin} shows that every integral of this form
may be reduced to dilogarithms, albeit with
the possibility of complex arguments.
Pursuing the methods of~\cite{Lewin}, one readily
establishes that $C(a,b)$ is reducible to real dilogarithms for
$a^2+b^2>4$. Implementing the algorithm of~\cite{Lewin}, in {\sc Reduce},
we obtained a formidably complicated result, involving 2 square roots:
$\sqrt{a^2+b^2-4}$ and $\sqrt{2b(b+2)}$. The appearance of the former is
to be expected; the characteristics of the result clearly change
when $a^2+b^2-4$ changes sign. The appearance of the latter
is a gratuitous consequence of the dispersive derivation; it may be
removed by consideration of $C(b,a)=C(a,b)$, but then $\sqrt{2a(a+2)}$
appears. Clearly there must exist a result involving neither
$\sqrt{2a(a+2)}$ nor $\sqrt{2b(b+2)}$. How to achieve this is problematic.

One strategy for removing a bogus square root is to differentiate
the dilogarithms that involve it and then to combine the resultant
logarithms, to show that the differential is free of the unwanted square
root. In this case, {\sc Reduce} showed that the dispersion
relation for the Feynman tetrahedron $C(a,b)$ yields the partial differential
equation
\begin{eqnarray}
\frac{b\sqrt{a^2+b^2-4}}{4}\,\frac{\partial}{\partial a}\,
\frac{a\sqrt{a^2+b^2-4}}{4}\,C(a,b)&=&\log\left(\frac{a+2}{a+b+2}\right)
+\frac{b}{a+2}\log\left(\frac{a+b+2}{b+2}\right)\nonumber\\&&{}
+\frac{2b}{a^2-4}\log\left(\frac{a+2}{4}\right),\label{pde}
\end{eqnarray}
which entails only the physical square root, easily traceable
to a tree diagram for elastic scattering~\cite{CTet}.
A second partial differential equation immediately
follows from the symmetry $C(a,b)=C(b,a)$ of the diagram.
We checked that the pair agrees with results in~\cite{AKR},
obtained by the methods of~\cite{AVK}, more recently espoused
in~\cite{ER}.

Systematic re-integration of~(\ref{pde}), by the methods of~\cite{Lewin},
still produced 20 dilogarithms, with 8 of these entailing the
unwanted square root $\sqrt{2b(b+2)}$. Accordingly,
we resorted to an alternative strategy, by evaluating $C(a,b)$
numerically at an arbitrarily chosen transcendental point,
$a=\exp(1),~b=\pi$, and then using the lattice algorithm PSLQ~\cite{PSLQ}
to search for a rational linear combination of dilogarithms of a character
suggested by those parts of the analytical 20-dilogarithm result
that did not involve the bogus square root $\sqrt{2b(b+2)}$.
After much trial and error, in search spaces of dimensions
as large as 80, to accommodate the possibility of many products of logs,
we found a simple log-free fit to the single numerical datum:
\begin{eqnarray}
\df{1}{8}abc\,C(a,b)&=&
{\rm Li}_2\left(-\frac{p}{m}\right)+{\rm Li}_2\left(1-\frac{4}{m}\right)
+{\rm Li}_2\left(1-\frac{m}{a+2}\right)
+{\rm Li}_2\left(1-\frac{m}{b+2}\right)\nonumber\\
&-&{\rm Li}_2\left(-\frac{m}{p}\right)-{\rm Li}_2\left(1-\frac{4}{p}\right)
-{\rm Li}_2\left(1-\frac{p}{a+2}\right)
-{\rm Li}_2\left(1-\frac{p}{b+2}\right)\label{dans}
\end{eqnarray}
with a dilogarithm ${\rm Li}_2(x):=-\int_0^x(dy/y)\log(1-y)$ and
\begin{equation}
c:=\sqrt{a^2+b^2-4}\,,\quad p:=a+b+2+c\,,\quad m:=a+b+2-c\,.
\end{equation}
Ansatz~(\ref{dans}) is manifestly symmetric in $(a,b)$
and fits the datum to 360-digit precision.

It was then a routine application of computer algebra
to prove that~(\ref{dans}) is correct,
by showing that it satisfies the partial differential
equation~(\ref{pde}).
Hence the r.h.s.\ of~(\ref{dans}) may differ
from the required result only by a function of $b$. But by symmetry
it thus differs only by a function of $a$, and hence only by a constant.
Since the r.h.s.\ and l.h.s.\ both vanish when $c=0$, the constant
must vanish. Hence~(\ref{dans}) is proven to be correct, though
no analytical {\em derivation} of it has yet been obtained.
To our knowledge, this is the first time that a lattice algorithm, such
as PSLQ, has been used to find a previously unknown solution to a pair
of partial differential equations.

We note that one of the 8 dilogarithms in~(\ref{dans}) may be removed,
using~\cite{Lewin}
\begin{equation}
0={\rm Li}_2(-p/m)+{\rm Li}_2(-m/p)+\df16\pi^2+\df12\log^2(p/m)\,.
\end{equation}
No further reduction was found by PSLQ, with transcendental values of $a$
and $b$. With rational values of $\{a,b,c\}$, considerable
simplification was obtained. For example
\begin{equation}
224\,C(14,8)={\rm Li}_2\left(\df35\right)+\df{1}{12}\pi^2-\log5\log\df95
\end{equation}
was spectacularly reduced by PSLQ to a single dilogarithm.
It remains an open question whether~(\ref{dans}) may be reduced to fewer
than 7 dilogs, in the general case. We suspect not.

\subsection{Reduction to Clausen values}

The result~(\ref{dans}) clearly entails only real dilogarithms
when $a^2+b^2>4$. When $a^2+b^2<4$, it may be reduced, by application
of Eq~(A.2.5.1) of~\cite{Lewin}, to Clausen values of the form
\begin{equation}
{\rm Cl}_2(\theta):=-\int_0^\theta d\phi\log\left|2\sin\df12\phi\right|
=\sum_{n>0}\frac{\sin(n\theta)}{n^2}\,.
\end{equation}
Since the imaginary part of a dilog yields 3 Clausen values, plus
the product of an angle and log, the result~(\ref{dans})
might be expected to be rather complicated, involving up to 16 terms.
Transforming to the regime where $\gamma:=\sqrt{4-a^2-b^2}$ is real,
one finds that
\begin{eqnarray}
\df{1}{16}ab\gamma\,C(a,b)&=&\df12\left\{
 {\rm Cl}_2(4\phi)
+{\rm Cl}_2(2\phi_a+2\phi_b-2\phi)
+{\rm Cl}_2(2\phi_a-2\phi)
+{\rm Cl}_2(2\phi_b-2\phi)\right.\nonumber\\&&\left.{}
-{\rm Cl}_2(2\phi_a+2\phi_b-4\phi)
-{\rm Cl}_2(2\phi_a)
-{\rm Cl}_2(2\phi_b)
-{\rm Cl}_2(2\phi)\right\}\label{cla}
\end{eqnarray}
is log-free and involves only 8 Clausen values, with arguments formed from
\begin{equation}
\phi:=\arctan\frac{\gamma}{a+b+2}\,,\quad
\phi_a:=\arctan\frac{\gamma}{a}\,,\quad
\phi_b:=\arctan\frac{\gamma}{b}\,,\label{phi}
\end{equation}
which are related by
\begin{equation}
\cos\phi_a\cos\phi_b
=\cos(\phi_a+\phi_b-2\phi)\,.
\label{key}
\end{equation}
The freedom from logs is highly non-trivial, entailing the
multiplicative relation
\begin{equation}
\left(1-\frac{4}{m}\right)\left(1-\frac{4}{p}\right)=
\left(1-\frac{m}{a+2}\right)\left(1-\frac{m}{b+2}\right)
\left(1-\frac{p}{a+2}\right)\left(1-\frac{p}{b+2}\right)
\end{equation}
between 6 of the arguments of the 8 dilogarithms of~(\ref{dans}).
Had it been known in advance that neither~(\ref{dans}) nor~(\ref{cla})
entails logs, while each reduces to only 8 terms, the process of constructing
a viable symmetric Ansatz would have been greatly simplified. We offer this
observation as a guide to future work.

\subsection{The symmetric tetrahedron}

To obtain a result for $C(1,1)$, we use the specific values
of the angles~(\ref{phi}), namely
$\phi=\alpha$, $\phi_a=\phi_b=\frac14\pi+\frac12\alpha$, with
$\alpha:=\arcsin\frac13$ appearing as the only non-trivial angle,
by virtue of~(\ref{class}). Then using ${\rm Cl}_2(\pi)=0$,
and the general identity~\cite{Lewin}
\begin{equation}
\df12{\rm Cl}_2(\pi-2\alpha)
={\rm Cl}_2(\df12\pi-\alpha)
-{\rm Cl}_2(\df12\pi+\alpha)\,,
\end{equation}
one finds that only the first and last terms in~(\ref{cla}) survive, giving
\begin{equation}
\frac{C(1,1)}{8\sqrt2}=\df12\left\{{\rm Cl}_2(4\alpha)-{\rm Cl}_2(2\alpha)
\right\}\approx0.01537\,,
\label{sym}
\end{equation}
in agreement with~(\ref{ans}). The tiny value will be seen to be significant.

\section{Connection to 1-loop diagrams}

In~\cite{DD}, Andrei Davydychev and Bob Delbourgo considered an apparently
very different problem, namely the massive 1-loop box diagram in 4 dimensions,
which yields a result uncannily similar to~(\ref{dans},\ref{cla}), in the
case of a common mass on the internal lines and a common norm
for the external 4-momenta. Then there are three kinematic
variables, which may be taken as Mandelstam's $\{s,t,u\}$.
The internal mass provides the scale, here set to unity.
In certain kinematic regimes, $\{s,t,u\}$ may be transformed to the
3 non-trivial dihedral angles, $\{\psi_1,\psi_2\,\psi_3\}$, of a
bi-rectangular tetrahedron in a 3-space of constant curvature~\cite{DD}.
This is one of the 4 congruent parts that result from
dissection of a tetrahedron with a symmetry that derives from the
common internal mass. The result then entails its volume, which is
a Schl\"afli~\cite{Schl,Cox} function.

After the results~(\ref{dans}) and~(\ref{cla}), for the
3-loop vacuum diagram, were communicated to Andrei Davydychev,
he made the intriguing suggestion that~(\ref{cla}), for the case $a^2+b^2<4$,
might be reducible from 8 real Clausen values to 7,
as is the case~\cite{DD} for the box diagram, in restricted kinematic
regimes. If this were the case, one might hope to cap the `magic'
feat in~\cite{magic}, where a 2-loop vacuum diagram was transformed
to a massless 1-loop triangle diagram, in a dimension differing by 2 units.
In the present case, such a conjuring act would entail
a more remarkable connection,
between diagrams whose loop numbers differ by 2,
while their spacetime dimensions differ only by unity.
We now examine this issue.

\subsection{Geometric and non-geometric boxes}

{}From~\cite{DD}, we obtained a simple conversion of
$\{s,t,u\}$ to $\{\psi_1,\psi_2,\psi_3\}$
as follows. Let
\begin{equation}
v:={4\over s}\,,\quad w:={4\over t}\,,\quad
x:=\frac{8}{s+t+u-8}\label{kin}
\end{equation}
be a re-parametrization of Mandelstam space.
Then the dihedral angles satisfy
\begin{equation}
{1-w\over\tan^2\psi_1}={\tan^2\psi_2\over1-x^2}={1-v\over\tan^2\psi_3}
={1\over\tan^2\delta}=G:={vw\over x^2}-(1-v)(1-w)
\label{psi}
\end{equation}
where $G$ derives from a Gram determinant and
$\delta$ is an auxiliary angle, with
\begin{equation}
\tan\delta\cos\psi_1\cos\psi_3
=D(\psi_1,\psi_2,\psi_3):=
\sqrt{\cos^2\psi_2-\sin^2\psi_1\sin^2\psi_3}\,.\label{delta}
\end{equation}
The box diagram evaluates to
\begin{equation}
B(s,t,u):=\frac{N(\psi_1,\psi_2,\psi_3)}{D(\psi_1,\psi_2,\psi_3)}\,,
\label{box}
\end{equation}
with a numerator that is a Schl\"afli function~\cite{DD}:
\begin{eqnarray}
N(\psi_1,\psi_2,\psi_3)&:=&\df12\left\{
 {\rm Cl}_2(2\psi_1+2\delta)-{\rm Cl}_2(2\psi_1-2\delta)
+{\rm Cl}_2(2\psi_3+2\delta)-{\rm Cl}_2(2\psi_3-2\delta)
\right.\nonumber\\&&\left.{}
-{\rm Cl}_2(\pi-2\psi_2+2\delta)+{\rm Cl}_2(\pi-2\psi_2-2\delta)
+2{\rm Cl}_2(\pi-2\delta)\right\}\,.\label{schl}
\end{eqnarray}
When $\{1-v,1-w,1-x^2,G\}$ are all positive,
$\{\psi_1,\psi_2,\psi_3,\delta\}$ are all real and~(\ref{schl})
is 4 times the volume of a bi-rectangular tetrahedron in
hyperbolic space, since the full tetrahedron may be dissected into
4 congruent bi-rectangular parts~\cite{DD}.

In the case that $\{1-v,1-w,1-x^2,G\}$ are all negative,
$\delta$ is imaginary, while
$\{\psi_1,\psi_2,\psi_2\}$ are real. Then both the numerator and
denominator of~(\ref{box}) are pure imaginary and we obtain a
geometric interpretation that entails the volume of
a tetrahedron in spherical space. For the residual sign possibilities,
there is {\em no\/}
interpretation in terms of real geometry. Indeed, unitarity
often requires the amplitude to be complex. Thus
vanishing of the Gram determinant of the external
momenta, at $G=0$, is emphatically {\em not\/} the signal for the geometry
to change from one sign of curvature to the other.
If one has a real geometry at some point $\{s,t,u\}$ near $G=0$,
then there is no real geometry at a neighbouring point,
with the opposite sign of $G$, since there~(\ref{psi})
forces $\{\psi_1,\psi_2,\psi_3\}$ to be imaginary.

By way of examples of geometric and non-geometric
behaviour, we consider $B_0(s,t):=B(s,t,-s-t)$, with light-like
external momenta. In the hyperbolic regime, we obtain
\begin{equation}
B_0(4,4)=4{\rm Cl}_2(\df12\pi)\,,\quad
B_0(6,6)=\frac{5{\rm Cl}_2(\df13\pi)}{\sqrt3}\,,\quad
B_0(\df{16}{3},\df{16}{3})=\frac{3{\rm Cl}_2(2\alpha)
+6{\rm Cl}_2(\df12\pi-\alpha)}{2\sqrt2}\label{al}
\end{equation}
with the first example giving 4 times Catalan's constant,
while the second is a rational multiple of a constant found in the 2-loop
4-dimensional vacuum diagram with 3 equal masses~\cite{BV}, which enjoys
a `magic' connection~\cite{magic} to a massless 1-loop triangle diagram.
The final example entails $\alpha:=\arcsin\frac13$, though in a
manner markedly different from~(\ref{ans}).

Non-geometric results
are obtainable from the instructive duality relation
\begin{equation}
B_0(s,t)-B_0(\lambda/s,\lambda/t)
=\frac{2}{\sqrt{\lambda}}\,\arccos\left({s\over2}-1\right){\rm arccosh}
\left({\lambda\over2s}-1\right)
+\left\{s\leftrightarrow t \right\}\,,\label{dual}
\end{equation}
with $\lambda:=4s+4t-st$.
It was proven by analytic continuation of~(\ref{schl}), after the
discovery by PSLQ that
\begin{equation}
B_0(\df83,\df83)-B_0(\df{16}{3},\df{16}{3})
=\frac{3(\df12\pi-\alpha)\log3}{2\sqrt2}\,,
\end{equation}
with product terms familiar from~\cite{CTet,sixth,poly}.
When one box in~(\ref{dual}) is geometric,
the products of angles and logs show that its dual is not.
Since~(\ref{cla}) has no such product, it cannot be such a
non-geometric box. We now consider whether it might be geometric.

\subsection{Obstacles to a single 3-loop vacuum volume}

Analytical considerations and numerical investigations, alike,
suggest that no geometric interpretation as a single tetrahedral volume,
and hence no relation to a single 1-loop diagram, is obtainable for
the 3-loop 3-dimensional vacuum diagram $C(a,b)$.

The argument against a real tetrahedral volume in spherical space
is compelling: the formula for such a volume
involves the real parts of complex ${\rm Li}_2$ values~\cite{DD,Cox}.
In contrast, our result~(\ref{dans}) entails purely real ${\rm Li}_2$
values when $a^2+b^2>4$. Thus the simplicity of the vacuum
diagram seems to preclude a geometric interpretation
in a space of positive curvature, since any such interpretation
would be too complicated, analytically speaking.

We argue that there is no interpretation as a single volume
in hyperbolic space, for $a^2+b^2<4$. Here we are guided by the fact that
all attempts to reduce the Clausen values in~(\ref{cla}) from 8 to 7,
as would be required by~(\ref{schl}), met with abject failure.

Since no-go claims based on analysis are notoriously fallible,
we also investigated the situation empirically, using PSLQ.
The first step was clear: is there a simple integer relations
between the 8 Clausen values in~(\ref{cla})? PSLQ replied with an
emphatic {\em no}, by proving that any integer
relation would entail a coefficient in excess of $10^{30}$.

Then we considered relations between
Clausen values generated by Abel's identity for 5 dilogarithms~\cite{Lewin}.
Since the imaginary part of a dilogarithm generates 3 Clausen values,
the generic relation will entail 15 Clausen values. The symmetric form
of the result is
\begin{equation}
0=\sum_{6\ge k\ge1}\theta_k=\sum_{6\ge k\ge1}\sin\theta_k
\quad\Longrightarrow\quad
0=\sum_{6\ge j>k\ge1}{\rm Cl}_2(\theta_j+\theta_k)\,,\label{15}
\end{equation}
with 6 angles, whose values and sines sum to zero, producing 15 Clausen
values, which also sum to zero.
{}From~(\ref{key},\ref{15}) we derived 3
relations
between Clausen values whose arguments are linear combinations of
$\{\phi,\phi_a,\phi_b\}$. A pair is formed by
\begin{eqnarray}
0&=&2{\rm Cl}_2(2\phi)-4{\rm Cl}_2(2\phi_b)+{\rm Cl}_2(4\phi_b)
+2{\rm Cl}_2(2\phi_b-2\phi)-2{\rm Cl}_2(2\phi_a-2\phi)
\nonumber\\&&{}+{\rm Cl}_2(2\phi_a-4\phi)+2{\rm Cl}_2(2\phi_a+2\phi_b-2\phi)
-{\rm Cl}_2(2\phi_a+4\phi_b-4\phi)\label{first}
\end{eqnarray}
and its $a\leftrightarrow b$ transform, while the third is symmetric:
\begin{eqnarray}
0&=&2{\rm Cl}_2(2\phi)-2{\rm Cl}_2(2\phi_a-2\phi)-2{\rm Cl}_2(2\phi_b-2\phi)
+{\rm Cl}_2(2\phi_a-4\phi)+{\rm Cl}_2(2\phi_b-4\phi)\nonumber\\&&{}
-2{\rm Cl}_2(2\phi_a+2\phi_b-2\phi)+2{\rm Cl}_2(2\phi_a+2\phi_b-4\phi)
-{\rm Cl}_2(4\phi_a+4\phi_b-8\phi)\nonumber\\&&{}
+{\rm Cl}_2(2\phi_a+4\phi_b-4\phi)+{\rm Cl}_2(4\phi_a+2\phi_b-4\phi)\,.
\label{third}
\end{eqnarray}

The next step was to engage PSLQ to search for more relations.
At the arbitrarily chosen transcendental point $a=\exp(-1)$, $b=1/\pi$,
we computed, to 360-digit precision, 44 Clausen values of the form
${\rm Cl}_2(2j\phi_a+2k\phi-2n\phi)$, with non-negative integers
bounded by $j<3$, $k<3$, $n<5$, $j+k+n>0$. PSLQ found only the
3 known relations. Moreover, it proved that any other relation
would involve an integer in excess of $10^5$.
Enlarging the search space to include angles in which the coefficients
of $\phi_a$ and $\phi_b$ differ in sign, we found no new relation.
It is easy to show that the 3 proven relations
do not enable a reduction of~(\ref{cla}) to less than 8 Clausen values.
Hence, a reduction to 7 real Clausen values,
as required for a single Schl\"afli function, would seem to require
a non-linear transformation of angles, for which we have seen no precedent.

\subsection{Reduction of diagrams to ideal tetrahedra}

The difficulty in relating~(\ref{cla}) to a geometric box
is more apparent when one writes it in terms of {\em differences\/} of
volumes of ideal hyperbolic tetrahedra.
An ideal tetrahedron has all its vertices at infinity
and is specified by 2 dihedral angles, $\theta_1$ and $\theta_2$,
at adjacent edges. The dihedral angle at the edge adjacent to these
is $\theta_3:=\pi-\theta_1-\theta_2$. Each remaining edge has
a dihedral angle equal to that at its opposite edge.
The volume of such a ideal tetrahedron is~\cite{DD}
\begin{equation}
V(\theta_1,\theta_2):=\df12\sum_{k=1}^3{\rm Cl}_2(\theta_k)
=\df12\{{\rm Cl}_2(2\theta_1)
+{\rm Cl}_2(2\theta_1)-{\rm Cl}_2(2\theta_1+2\theta_2)\}\,.\label{ideal}
\end{equation}
Thus~(\ref{cla}) may be written, rather neatly, as
\begin{equation}
\df{1}{16}ab\gamma\,C(a,b)
=V(\phi,\psi_a)+V(\phi,\psi_b)
-V(\phi,\psi_a+\psi_b)-V(\phi,\phi)\,,\label{neat}
\end{equation}
where $\psi_{a,b}:=\phi_{a,b}-\phi$ are confined to the
interval $[\phi,\pi/2-\phi]$, with $\phi$ confined to $[0,\pi/4]$.
Similarly, the box volume~(\ref{schl}) may be written as
\begin{eqnarray}
N(\psi_1,\psi_2,\psi_3)
&=&V(\delta+\psi_1,\delta-\psi_1)+V(\delta+\psi_3,\delta-\psi_3)\nonumber\\
&+&V(\df12\pi+\psi_2-\delta,\df12\pi-\psi_2-\delta)
+V(\df12\pi-\delta,\df12\pi-\delta)\label{Nis}
\end{eqnarray}
with the auxiliary angle $\delta$ given by~(\ref{delta}).

Both the vacuum result~(\ref{neat}) and the box volume~(\ref{Nis})
are non-negative, in their hyperbolic regimes, where the angles are real.
Now we consider their zeros and maximum
values. The box volume~(\ref{Nis}) vanishes only for
$\cos\psi_2=\sin\psi_1\sin\psi_3$, where the denominator~(\ref{delta})
of the diagram vanishes, at the boundary of the hyperbolic regime.
The maximum volume is $N(0,0,0)=4{\rm Cl}_2(\pi/2)$,
achieved in the case of the box diagram $B_0(4,4)$ in~(\ref{al}),
with $D(0,0,0)=1$.
In contrast, the vacuum diagram yields a combination~(\ref{neat}) of
ideal tetrahedral volumes that vanishes at $b=0$,
where $\psi_a=\phi$ and $\psi_b=\frac12\pi-\phi$, with
the last term cancelling the first,
and the third cancelling the second; and at $a=0$,
with the last cancelling the
second, and the third cancelling the first; and at $\gamma=0$,
where all terms vanish separately. Its maximum value occurs at the
totally symmetric point $a=b=1$, where~(\ref{sym}) gives
a combination of volumes that is more than 200 times smaller than the
maximum volume, $N(0,0,0)=3.66386237$, achieved by the box.

{}From the above, the difficulty of relating the vacuum diagram to
a box is glaring. The geometric insight of~\cite{DD} led to the
conclusion that every 4-dimensional 1-loop box diagram
may be evaluated by dissecting\footnote{We discount the possibility
that this dissection might entail subtraction of
bi-rectangular volumes in the totally symmetric case~(\ref{sym}).}
its associated volume into
no more than 6 bi-rectangular parts, each given by a Schl\"afli
function. We have shown that the addition and subtraction
of ideal tetrahedra, entailed by the vacuum diagram in~(\ref{neat}),
leads to net volumes that are, typically, two orders of magnitude smaller
than the volumes associated with a box diagram, via the additions
in~(\ref{Nis}).

Yet there {\em is\/} a remarkably strong connection between
3-loop vacuum diagrams and 1-loop boxes: both entail
{\em combinations\/} of volumes of ideal tetrahedra.
We have show this for the vacuum diagram~(\ref{ideal}).
In the more complicated case of an arbitrary box diagram, one may
obtain up to 24 ideal tetrahedra, with each of the 6 bi-rectangular
constituents~\cite{DD} of a general tetrahedron yielding
4 ideal tetrahedra, via~(\ref{Nis}).
Moreover every such ideal tetrahedron equates to
a massless 1-loop triangle diagram~\cite{DD,magic}.

We conclude that 3-loop 3-dimensional vacuum diagrams and
1-loop 4-dimensional boxes do not equate, directly.
Rather, they share a common reduction, via hyperbolic geometry,
to 1-loop massless 4-dimensional triangle diagrams, i.e.\ ideal tetrahedra.

\section{Hyperbolic manifolds from multi-loop diagrams}

The box-diagram value $B_0(4,4)=N(0,0,0)=4{\rm Cl}_2(\pi/2)=
3.66386237$ is familiar in an apparently quite different context:
it is the hyperbolic volume complementary to Whitehead's
2-component link, with 5 crossings~\cite{Adams}. Like the
majority of knots and links, this link is hyperbolic,
which means that the 3-manifold complementary to it admits a
metric of constant negative curvature. The volume of this
hyperbolic manifold is then an invariant~\cite{AHW} associated with the link.
The Borromean rings have a volume twice as large, namely 8 times Catalan's
constant. Moreover, the numerator $N(\pi/4,0,\pi/4)=\frac52{\rm Cl}_2(\pi/3)$
of the box diagram $B_0(6,6)$ in~(\ref{al}) is a rational
multiple of the volume, $2{\rm Cl}_2(\pi/3)=2.02988231$,
of the figure-8 knot, which is the unique knot with 4 crossings.

The common analytical feature of such link invariants
and the Feynman diagrams of this paper is the volume,~(\ref{ideal}),
of an ideal tetrahedron. It may be regarded as
a real-valued function of a single complex variable~\cite{AMS}:
\begin{equation}
{\cal V}(z):=\Im\left\{{\rm Li}_2(z)
+\log\left|z\right|\log(1-z)\right\}=V(\arg(z),-\arg(1-z))
\label{vz}
\end{equation}
with dihedral angles that are the arguments of $\{z,1/(1-z),1-1/z\}$.
The symmetries
\begin{equation}
{\cal V}(1-z)={\cal V}(1/z)={\cal V}(\overline{z})=-V(z)\,,\label{12}
\end{equation}
where $\overline{z}$ is the complex conjugate of $z$, imply that
$\{z,1/\overline{z},1/(1-z),
1-\overline{z},1-1/z,\overline{z}/(\overline{z}-1)\}$
all give the same value for~(\ref{vz}), while their conjugates give
a result differing only in sign. The hyperbolic volume of a knot or
link is expressible as a finite number of ideal terms
of the form~(\ref{vz}),
with arguments that result from complex roots of
polynomials~\cite{Adams,AHW}.
For example, the volume of the figure-8 knot is $2{\cal V}(z)$, with
$z(1-z)=1$, while the volume of the Borromean rings is
$8{\cal V}(z)$, with $z(1-z)=\frac12$.
The sole hyperbolic 5-crossing knot, $5_2$,
has a volume, 2.8281220, given by $3{\cal V}(z)$, with $z^3=z-1$.
This cubic gives the relation $3\theta_1+\theta_2=\pi$ between
the dihedral angles in~(\ref{vz}). We shall meet it again, at 12 crossings,
in the context of 8-loop quantum field theory.

\subsection{Positive hyperbolic knots at 7 loops}

In~\cite{BK15}, Dirk Kreimer and I considered knots with up to
15 crossings, classifying the numerical content
of field-theory counterterms up to 9 loops. An account of the wider
issues is provided by~\cite{DK}.
The knots in question are all
positive, i.e.\ their minimal braidwords involve only positive
powers of the generators, $\sigma_k$, of the braid group~\cite{VJ}.
A consequence is that
no hyperbolic knot is encountered in the analysis of diagrams with
less than 7 loops, where only torus knots are encountered.
At the 7-loop level one encounters two 10-crossing knots
that are both positive and hyperbolic, with braidwords
$10_{139}=
\sigma_1^{}\sigma_2^{3}\sigma_1^{3}\sigma_2^{3}$
and $10_{152}=
\sigma_1^{2}\sigma_2^{2}\sigma_1^{3}\sigma_2^{3}$,
offering the first possibility to study the reduction
to ideal tetrahedra of knots implicated by counterterms.
Numerical triangulations were obtained, at 12-digit accuracy,
from Jeff Weeks' program SnapPea~\cite{snap}. We then used PSLQ to
identify the relevant polynomials, whose roots were extracted
to 50 digits, giving
\begin{eqnarray}
V_{10_{139}}&=&{\tt
4.85117075733273756705832705211531247884528302776999
}\label{139}\\
V_{10_{152}}&=&{\tt
8.53606534720560860314418192054932599496499139691401
}\label{152}
\end{eqnarray}
as the volumes of the positive 10-crossing knots.
SnapPea identified the manifold complementary to $10_{139}$ as isometric to
entry m389 in its census. Its volume coincides with that of
m391, for the 8-crossing 2-component link labelled $8_2^2$
in the appendices of~\cite{Adams} and~\cite{Rolfsen}.
The manifold complementary to $10_{152}$
has a volume greater than any in SnapPea's census
of 6,075 cusped manifolds triangulated by not more than 7 tetrahedra.

We found that~(\ref{139}) results from a remarkably simple triangulation,
\begin{equation}
V_{10_{139}}=4{\cal V}(z)+{\cal V}(z^2+1)\,;\quad z^2+1=z^2(z-1)^2\,,
\label{139is}
\end{equation}
with a matching condition that makes the dihedral
angles of the second term linear combinations of those of the first.
This simplicity is in marked contrast to
\begin{eqnarray}
V_{10_{152}}&=&{\cal V}(z)+2{\cal V}(z+1)
+{\cal V}(2z-z^2)+{\cal V}(z^2(z-1)^2)
+4{\cal V}\left(\frac{2z-1}{2z-z^2}\right)
\,;\nonumber\\&&{}\quad
z(2z-z^2)^2=(z+1)^2(z-1)\,,\label{152is}
\end{eqnarray}
whose quintic produces 15 distinct Clausen values, with angles
reducible to linear combinations of the arguments of
$\{z^2,(z+1)^2,(z-1)^2,(2z-1)^2\}$.
The simpler form of~(\ref{139is}),
with only 2 distinct tetrahedra, and only 2 linearly independent angles,
accords with the experience of~\cite{BKP}, where it
was found that $10_{139}$ is simpler than $10_{152}$ in the field-theory
context, since it is more readily obtained from the skeining of
link diagrams that encode the intertwining of momenta in 7-loop diagrams.

\subsection{Positive hyperbolic knots at 8 loops}

Observing the contrasting reductions~(\ref{139is},\ref{152is})
to ideal volumes, we proceeded to 12 crossings, relevant to 8-loop
counterterms~\cite{BK15}. Work with John Gracey
and Dirk Kreimer~\cite{BGK} had focussed on a pair of positive hyperbolic
knots, $12_A:=\sigma_1^{}\sigma_2^{7}\sigma_1^{}\sigma_2^{3}$ and
$12_B:=\sigma_1^{}\sigma_2^{5}\sigma_1^{}\sigma_2^{5}$,
one of which is associated with the appearance of the
irreducible~\cite{DZ,BBG,Eul} double Euler sum
$\zeta_{9,3}:=\sum_{j>k>0}j^{-9}k^{-3}$
in counterterms, while the other relates to a quadruple sum that cannot be
reduced to simpler non-alternating sums, and was found in~\cite{Eul} to
entail the alternating Euler sum
$U_{9,3}:=\sum_{j>k>0}(-1)^{j+k}j^{-9}k^{-3}$.
In~\cite{BGK} we tentatively identified $12_A$ as the knot associated with
$\zeta_{9,3}$, by study of counterterms in the large-$N$ limit, at $O(1/N^3)$,
where $\zeta_{9,3}$ occurs, but $U_{9,3}$ is absent.
Thus we expect $12_A$ to have a
simpler reduction to ideal tetrahedra than that for $12_B$.

This expectation was notably confirmed by computation, which
gave the volumes
\begin{eqnarray}
V_{12_A}&=&{\tt
2.82812208833078316276389880927663494277098131730065
}\label{12a}\\
V_{12_B}&=&{\tt
5.91674573518278869527226015189683245321707317046868
}\label{12b}
\end{eqnarray}
with triangulations that SnapPea identified with the manifolds m016 and v2642.
The result for $12_A:=\sigma_1^{}\sigma_2^{7}\sigma_1^{}\sigma_2^{3}$
is indeed rather special: the volume is equal to that of manifold m015,
for the hyperbolic knot with 5 crossings\footnote{Section~5.3
of~\cite{Adams} gives an excellent introduction to hyperbolic knots.
Unfortunately, Fig~5.29 is misdrawn, depicting
$\sigma_1^{}\sigma_2^{-7}\sigma_1^{}\sigma_2^{3}$, with manifold v0960,
instead of the positive knot
$12_A:=\sigma_1^{}\sigma_2^{7}\sigma_1^{}\sigma_2^{3}$.}:
\begin{equation}
V_{12_A}=V_{5_2}=3{\cal V}(z)\,;\quad z^3=z-1\,.\label{12ais}
\end{equation}
Equalities between volumes of hyperbolic knots are rare, with none occurring
at less than 10 crossings.
It is intriguing that the knot $12_A$,
identified with $\zeta_{9,3}$ in~\cite{BGK},
has a triangulation as simple as that for the knot $5_2$.
By contrast the
result for $12_B:=\sigma_1^{}\sigma_2^{5}\sigma_1^{}\sigma_2^{5}$,
\begin{eqnarray}
V_{12_B}&=&4V(\psi_1,\psi_2)
+V(2\psi_1,2\psi_1+2\psi_2)
+2V(3\psi_1+\psi_2,\psi_1-\psi_2)\,;\nonumber\\&&
\psi_1=\arg(z)\,;\quad\psi_2=-\arg(1-z)\,;\quad4z^4=2z^2-2z+1\,,
\label{12bis}
\end{eqnarray}
involves 9 distinct Clausen values, with angles coming from
the solution to a quartic. As before, the relative ease with which
positive hyperbolic knots are obtained from Feynman diagrams is reflected by
the relative simplicity of their triangulations.
As further confirmation of this trend, we cite the cases of the remaining 5
positive knots with 12 crossings, which were not obtained from skeining
counterterms in~\cite{BGK}, nor related to Euler sums in~\cite{BK15}.
Their volumes exceed that of~(\ref{12b}), ranging from 7.40 to 13.64,
with commensurately complicated triangulations.

It thus appears that
Feynman diagrams entail positive knots that are either not hyperbolic,
as in the case of torus knots, which suffice through 6 loops, or
`marginally' hyperbolic, with a small volume, related to a relatively
simple triangulation.

\subsection{A simple hyperbolic volume at infinite loops}

We now study the volume, $V_{2n}$, of the positive $2n$-crossing
knot $K_{2n}:=\sigma_1^{}\sigma_2^{2n-5}\sigma_1^{}\sigma_2^{3}$,
related to double Euler sums of weight $2n$
in counterterms at $n+2\ge6$ loops~\cite{BK15,BGK}.
We found that this volume is bounded, as $n\to\infty$.

Since $K_{8}=8_{19}$ and $K_{10}=10_{124}$
are the (4,3) and (5,3) torus knots, $V_{8}=V_{10}=0$. At 12 crossings,
$V_{12}:=V_{12_A}$ is given by~(\ref{12a},\ref{12ais});
the appendix of~\cite{Adams} shows that no hyperbolic knot from 6 through
9 crossings has a volume as small as this. We found that $K_{14}$,
with manifold m223, has the same volume, 4.12490325, as
$8_{20}=\sigma_1^{}\sigma_2^{-3}\sigma_1^{}\sigma_2^3$, with
manifold m222. In general, the volume of the $2n$-crossing positive
knot $K_{2n}:=\sigma_1^{}\sigma_2^{2n-5}\sigma_1^{}\sigma_2^{3}$,
with $2n\ge12$,
coincides with that of the non-positive knot
$\sigma_1^{}\sigma_2^{11-2n}\sigma_1^{}\sigma_2^{3}$,
formally obtained by $n\to8-n$, and hence having a crossing
number that cannot exceed $2n-6$.

The manifolds of $K_{16}$ and $K_{18}$ were identified
as s384 and v0959,
triangulated by 6 and 7 tetrahedra, respectively; their volumes are not much
larger than that of $K_{14}$. Moreover, the trend of
\begin{equation}
\begin{array}{ll}
V_{14}={\tt4.124903252}\qquad&V_{30}={\tt5.227842810}\\
V_{16}={\tt4.611961374}\qquad&V_{32}={\tt5.244429225}\\
V_{18}={\tt4.854663387}\qquad&V_{34}={\tt5.257409836}\\
V_{20}={\tt4.993271973}\qquad&V_{36}={\tt5.267755714}\\
V_{22}={\tt5.079718733}\qquad&V_{38}={\tt5.276132543}\\
V_{24}={\tt5.137154054}\qquad&V_{40}={\tt5.283008797}\\
V_{26}={\tt5.177195133}\qquad&V_{42}={\tt5.288721773}\\
V_{28}={\tt5.206190226}\qquad&V_{44}={\tt5.293519248}
\end{array}\label{Kvol}
\end{equation}
suggests an asymptotic value
\begin{eqnarray}
V_\infty&=&3{\rm Cl}_2(2\omega)-3{\rm Cl}_2(4\omega)+{\rm Cl}_2(6\omega)
\label{infty}\\&=&{\tt
5.33348956689811958159342492522130008819676777710528
}\nonumber
\end{eqnarray}
with $\omega:=\arctan\sqrt7$, which is equal to the volume
\begin{equation}
V_{(\sigma_1^{2}\sigma_2^{-1})^2}=4{\cal V}(z)+2{\cal V}(2z)\,;
\quad2z^2=3z-2\label{s776}
\end{equation}
of manifold s776, complementary to the 6-crossing 3-component link
$6_1^3:=(\sigma_1^{2}\sigma_2^{-1})^2$.
To test~(\ref{infty}), we used SnapPea to evaluate volumes for a
selection of crossing numbers from 50 up to 500,
corresponding to counterterms with up to 252 loops.
The tight bounds
\begin{equation}
(\df14n-1)^2\left\{V_\infty-V_{2n}\right\}\in[0.811,0.816]\,;
\quad2n\in[50,500]\,,\label{limits}
\end{equation}
make a compelling case for the asymptotic behaviour
\begin{equation}
V_{2n}=V_\infty-\frac{C}{(\frac14n-1)^2}+O(n^{-4}),
\label{asy}
\end{equation}
with an invariance under $n\to8-n$, noted above, and a constant
$C=0.8160\pm0.0001$.

Thus we come full circle, from an infinite number of loops
back to a 1-loop result, since~(\ref{infty}) relates directly
to a 1-loop box, with
\begin{equation}
V_\infty=V_{(\sigma_1^{2}\sigma_2^{-1})^2}
=3N(\df13\pi,0,\df13\pi)=\df34\sqrt7\,B_0(7,7)\label{b7}
\end{equation}
being 3 times the volume of the light-like
equal-mass box diagram of Section~3.1, at $s=t=7$. This complements
the link invariants obtained at $s=t=4$ and $s=t=6$ in~(\ref{al}).
Moreover, the 12-crossing 3-component
link $(\sigma_1^2\sigma_2^{-2})^3$ has a volume
\begin{eqnarray}
V_{(\sigma_1^2\sigma_2^{-2})^3}
&=&6N(\df16\pi,0,\df16\pi)=\df32\sqrt{15}\,B_0(5,5)
\label{b5}\\&=&{\tt
18.83168336678760750554026296116895115755581340126291
}\nonumber
\end{eqnarray}
which is 6 times the volume of the box diagram at $s=t=5$.
Thus we now have 4 relations between Feynman diagrams and link invariants.
\begin{enumerate}
\item
The volume of the figure-8 knot, $4_1$, is
$2{\rm Cl}_2(\pi/3)=\frac45N(\pi/4,0,\pi/4)$. This Clausen value occurs in
the 2-loop equal-mass vacuum diagram~\cite{BV}, the 1-loop massless triangle
diagram at its symmetric point~\cite{magic}, and the equal-mass
light-like box diagram of~\cite{DD} at $s=t=6$.
\item
The volumes of the Whitehead link, $5_1^2$,
and the Borromean rings, $6_2^3:=(\sigma_1^{}\sigma_2^{-1})^3$,
are multiples of Catalan's constant, ${\rm Cl}_2(\pi/2)=\frac14N(0,0,0)$.
This Clausen value results at the simultaneous threshold values $s=t=4$
of the box.
\item
At $s=t=5$ we obtain the volume of the link
$(\sigma_1^2\sigma_2^{-2})^3$
in~(\ref{b5}).
\item
At $s=t=7$ we obtain the volume of the link
$6_1^3:=(\sigma_1^{2}\sigma_2^{-1})^2$
in~(\ref{b7}).
This is also the infinite-loop limit of the hyperbolic volumes
of the knots $\sigma_1^{}\sigma_2^{2n-5}\sigma_1^{}\sigma_2^{3}$,
associated in~\cite{BK15,BGK} with the appearance in
counterterms~\cite{DK,BKP}, at $n+2\ge6$ loops, of irreducible double
Euler sums~\cite{Eul,BBBL} of weight $2n$.
\end{enumerate}

There are further cases of knots and links whose volumes
entail a single Schl\"afli function, and hence a single box diagram.
Harnessing PSLQ to SnapPea, we obtained
\begin{eqnarray}
V_{9_{41}}&=&10N(\df25\pi,\df1{10}\pi,\df15\pi)=
10{\rm Cl}_2(\df25\pi)+5{\rm Cl}_2(\df45\pi)\label{9_41}\\&=&{\tt
12.09893602599078738356455696387624160295557377848341
}\nonumber\\
V_{10_{123}}&=&10N(\df3{10}\pi,\df15\pi,\df1{10}\pi)=
15{\rm Cl}_2(\df25\pi)+5{\rm Cl}_2(\df45\pi)\label{10_123}\\&=&{\tt
17.08570948298286127690097484048365482503835960943063
}\nonumber
\end{eqnarray}
for the volumes of the knots $9_{41}$ and $10_{123}$, and
\begin{eqnarray}
V_{(\sigma_1^2\sigma_2^{-1})^3}&=&
6N(\df14\pi,\df16\pi,\df14\pi)\label{9_40^2}\\&=&{\tt
12.04609204009437764726837862923359423099605804944500
}\nonumber\\
V_{(\sigma_1^{}\sigma_2^{-2}\sigma_3^{}\sigma_2^{-2})^2}&=&
6N(\df16\pi,\df16\pi,\df16\pi)\label{4link}\\&=&{\tt
16.59129969483175048405984013396780188163367504042159
}\nonumber
\end{eqnarray}
for the 9-crossing 2-component link $9_{40}^2:=(\sigma_1^2\sigma_2^{-1})^3$
and the 12-crossing 4-component link
$(\sigma_1^{}\sigma_2^{-2}\sigma_3^{}\sigma_2^{-2})^2$.
At 8 crossings, we found that
\begin{eqnarray}
V_{8_{18}}&=&3{\rm Cl}_2(2\beta)+12{\rm Cl}_2(\df12\pi+\beta)
\label{8_18}\\&=&{\tt
12.35090620915820017473630443842615201419925670412000
}\nonumber\\
V_{8_{21}}&=&\df12V_\infty+\df13V_{8_{18}}
\label{8_21}\\&=&{\tt
6.783713519835126515708813942086034048831469456592638
}\nonumber
\end{eqnarray}
with $\beta:=\arcsin\frac{\sqrt2-1}{2}$.
Integer relations between volumes, as in~(\ref{8_21}),
appear to be fairly common; we cite $V_{7^2_6}=V_\infty+V_{5^2_1}$
as another example, with the infinite-loop limit
of the knots of~\cite{BK15} here appearing as the difference in volume
of a pair of 2-component links.

\section{Conclusions}

The volumes of ideal hyperbolic tetrahedra play (at least) 6 roles in
field theory.
\begin{enumerate}
\item They result from the evaluation of 3-loop 3-dimensional vacuum
diagrams, where their volumes tend to cancel, making the
maximum~\cite{CTet} value~(\ref{sym}) remarkably small.
\item They also result from 1-loop 4-dimensional box diagrams~\cite{DD},
where their volumes tend to add, giving $O(10^2)$ times the volume
of 3-loop vacuum diagrams.
\item Each ideal volume corresponds to a massless 1-loop
triangle diagram~\cite{DD}.
\item Each ideal volume also corresponds to a massive 2-loop
vacuum diagram~\cite{magic}.
\item The ease with which the volume of a positive hyperbolic knot is reduced
to ideal volumes is indicative of the ease with which the knot results from
skeining momentum flow in counterterms~\cite{DK,BKP}.
\item The family of knots $\sigma_1^{}\sigma_2^{2n-5}\sigma_1^{}\sigma_2^{3}$,
associated with multiple zeta values~\cite{DZ,Eul} in
counterterms~\cite{BK15,BGK} at $n+2\ge6$ loops, yields a hyperbolic
volume, at infinite loops, which is 3 times that for a simple 1-loop box.
\end{enumerate}

Conclusion~1 was obtained via~(\ref{dans}), for a
3-loop vacuum diagram, with 3 distinct masses, in 3 dimensions.
Its analytic continuation to the hyperbolic regime, $a^2+b^2<4$,
is given by~(\ref{cla}), which may expressed, as in~(\ref{neat}),
in terms of 4 volumes of ideal tetrahedra, 2 of which enter with minus
signs. Conclusions~2--4 result from the work in~\cite{DD,magic},
which we here extended by exposing the duality relation~(\ref{dual})
and showing how the additions in~(\ref{Nis}), for box diagrams,
tend to produce results
two orders of magnitude greater than those from the cancellations
in~(\ref{neat}), for 3-loop vacuum diagrams.
Conclusion~5 is based on contrasting~(\ref{139is})
with~(\ref{152is}), at 7 loops, and~(\ref{12ais}) with~(\ref{12bis}),
at 8 loops.
Conclusion~6 is based on the strong numerical evidence~(\ref{limits})
for the asymptote~(\ref{b7}), corresponding to the volume
of the link $6_1^3:=(\sigma_1^{2}\sigma_2^{-1})^2$,
which is 3 times that of the light-like equal-mass box diagram at $s=t=7$.

The discovery~(\ref{dans}), which sparked these hyperbolic connections,
is now proven, though it was not derived, in the traditional sense; instead
it was inferred by numerical investigation and then verified by routine
differentiation w.r.t.\ masses.
Similarly empirical methods led to~(\ref{dual},\ref{b7}).
Such procedures prompt a question: what is served by mathematical proof?
The result~(\ref{ans}) was discovered in~\cite{CTet} at modest numerical
precision, and then checked to 1,000 digits.
There was no shadow of doubt that it was correct, though unproven.
Now it is proven, yet by a method as thoroughly empirical as that
which enabled its discovery. More important than the proof itself
is the route to it, since discovery of~(\ref{dans}), with 3 distinct masses,
provides fertile ground for conjectures on behaviour with more mass scales,
or in 4 dimensions. A comparable situation was apparent in~\cite{sixth,poly},
where the results themselves, again from PSLQ, were more illuminating than
the {\em post hoc} proofs found for some of them. As Michael Atiyah has
remarked~\cite{MA}: if possession is nine tenths of the law,
discovery is nine tenths of the proof.

\noindent{\bf Acknowledgments:} I thank
David Bailey, for implementing PSLQ,
Andrei Davydychev, for suggesting a relation of~(\ref{dans}) to geometry,
Dirk Kreimer, for tuition in knot theory,
Al Manoharan, for converting SnapPea to Windows95,
Arttu Rajantie, for the stimulus to solve PDEs for vacuum diagrams, and
Don Zagier for stressing the importance of~(\ref{vz}).

\raggedright

\end{document}